\documentclass{article}
\usepackage{times,fancyhdr}
\usepackage{cite}
\usepackage{graphicx,amsmath,amssymb}

\title{NLO jet vertex from Lipatov's QCD effective action\footnote{Preprint numbers: LPN11-55, IFT-UAM/CSIC-11-73, FTUAM-11-57}} 
\author{Martin Hentschinski \& Agust{\' \i}n Sabio Vera \\ 
\\
Instituto de F{\' \i}sica Te{\' o}rica UAM/CSIC, Nicol{\'a}s Cabrera 15, \\ 
\& Universidad Aut{\' o}noma de Madrid, E-28049 Madrid, Spain} 

\begin{document} 


\maketitle 

Real and virtual contributions to the quark-initiated forward jet vertex in QCD are 
calculated at next-to-leading logarithmic accuracy using Lipatov's effective action, valid in the high 
energy multi-Regge limit. A regularization of longitudinal divergencies is proposed which allows for the determination of jet observables. Agreement with previous results in the literature is found.

\section{Introduction}

Simplifications in scattering amplitudes of Quantum Chromodynamics
(QCD) appear when the interaction takes place in the high energy Regge
limit of large center-of-mass energy. For inelastic processes a
generalization is to consider multi-Regge kinematics where the Regge
limit is applied to multiple sub-channels~\cite{BFKL1}. Virtual
contributions are then packed in a gluon Regge trajectory which
appears as a $s_i^{\omega(t_i)}$ factor in each sub-channel with
center-of-mass energy $\sqrt{s_i}$ and momentum transfer
$t_i$. Reggeized gluons then naturally emerge in the $t$-channel with
propagators indicating the no-emission probability of any particle in
the rapidity interval $\sim \log(s_i/|t_i|)$. At the edge of these
intervals emissions occur with a probability given by effective
vertices representing the interaction of reggeons with usual
particles.  The gluon Regge trajectory $\omega(t_i)$ has been
calculated at leading ${\cal O} \left({\alpha_s}\right)$ (LL) and
next-to-leading order ${\cal O} \left({\alpha_s^2}\right)$ (NLL) in
QCD and to all orders in ${\cal N}=4$ super Yang-Mills
theory~\cite{Trajectory}. At NLL, where terms of the form
$\left(\alpha_s \log{(s_i/t_i)} \right)^n$ and $\alpha_s
\left(\alpha_s \log{(s_i/t_i)} \right)^n$ are resummed, in each
production cluster one or two particles can be generated, this is the
so-called quasi-multi-Regge kinematics (QMRK)~\cite{BFKLNLO}. The
linearity of QMRK is broken at higher orders due to unitarity in all
channels, opening up transitions to multiple reggeized gluons in the
$t$-channel.

The effective vertices for the coupling of reggeons to usual particles
are complicated. An efficient tool to evaluate them is the high energy
effective action proposed by Lipatov~\cite{LevSeff}. It is based on
the QCD action with gauge fixing and ghost terms, with the addition of
an induced component written in terms of gauge-invariant currents
where Wilson lines generate color fields ordered in light cone
components, with a non-trivial interaction with reggeon fields. Some
transition vertices have been calculated using the Feynman rules
derived from this effective action~\cite{LevSeff} (see
also~\cite{Martin}).  In the present work the focus lies on obtaining
a vertex which is very important for phenomenological applications of
QMRK to hadron colliders~\cite{Vera:2006un}: the vertex for the
transition of a quark into a forward jet plus a remnant together with
an off-shell reggeon in the $t$-channel. At the large hadron collider
(LHC) at CERN there are several observables where QMRK should
dominate.  An example is that where jets are tagged in the forward
region, with a relatively large transverse momentum. When two of these
jets are each associated to one of the hadrons, and are at a large
relative rapidity with similar transverse momentum squared, then QMRK
applies. For the cross-section of this configuration the jet vertex
studied in this work is a crucial piece. It is also possible to make
the observable more exclusive by selecting events with a third jet in
the central regions of rapidity at the
detectors~\cite{Bartels:2006hg}.

In this letter a first section is provided with a discussion on
Lipatov's effective action together with the Feynman rules used in the
subsequent calculations. The bulk of the results lie in the sections
devoted to calculate the virtual and real corrections to the forward
jet vertex, with a discussion on previous results in the
literature. Finally, conclusions and suggestions for future work are
presented.

\section{High Energy Effective Action \& Feynman Rules}

The amplitudes of interest are the four-point amplitude with on-shell
external quarks of momenta $p_a+p_b \to p_1 +p_2$, and the five-point
amplitude with an extra on-shell gluon with momentum $q$ in the final
state. The squared center-of-mass energy is $s = (p_a+p_b)^2 = 2 p_a
\cdot p_b$. It is convenient to introduce the light-like four momenta
$n^\pm$ with $n^+ \cdot n^- = 2$ related to the incoming momenta by
$n^+ = 2 p_b / \sqrt{s}$ and $n^- = 2 p_a / \sqrt{s}$. The Sudakov
decomposition of a general vector $k^\mu$, using $k^\pm \equiv k \cdot
n^\pm$, is then $k^\mu = k^+ n^-/2+k^- n^+ /2 + k_\perp$, and $p_a =
p_a^+ n^- / 2, p_b = p_b^- n^+ /2$.

With these notations at hand it is now possible to give a brief
introduction to Lipatov's effective action, which contains two
pieces. The first one is the usual QCD action with gauge fixing and
ghost terms. The second one introduces new degrees of freedom, the
gauge invariant reggeized gluon fields $A_\pm (x) = - i t^a A_\pm^a
(x)$, with $t^a$ being the SU($N_c$) color matrices, and their
interactions with the gluon fields $v_\mu (x) = -i t^a v_\mu^a (x)$.
This induced part of the effective action generates vertices with
several reggeized gluons in the $t$-channel. To impose the condition
that all gluon emissions with momenta $q$ are ordered in longitudinal
components ($q^+_i \gg q^+_{i+1}, q^-_{i+1} \gg q_i^-$) the constraint
\begin{align}
  \label{eq:constraint}
  \partial_\pm A_\mp & = 0,  && \text{with}&    \partial_\pm &=
n_\pm^\mu \partial_\mu,
\end{align}
is always implied.  In QMRK clusters of particles are produced with a
large rapidity distance between them. Inside each cluster the induced
part of the effective action generating the interaction between
reggeized and usual fields reads
\begin{eqnarray}
 S_{\text{ind.}} [v_\mu, A_\pm]  = \int \text{d}^4 x {\rm Tr}\left[ \left(V_+ [v(x)] - A_+(x)\right)\partial_\mu^2
  A_-(x) + \left( V_- [v(x)]  - A_-(x)\right)\partial_\mu^2 A_+(x)\right]
\end{eqnarray}
where $V_\pm[v(x)] = - g^{-1} \partial_\pm {\rm P}\, {\rm exp}
\left(-\frac{g}{2} \int_{-\infty}^{x^\pm} dz^\pm v_\pm(z)\right)$
corresponds to an effective current formed by an array of fields
ordered in longitudinal components.  Note that the reggeized gluon
fields of the effective action $A_\pm(x)$ propagate in four space time
dimensions.  The Feynman rules derived from this action can be found
in~\cite{LevSeff}. The propagator for the reggeized gluon is
simple. There also exists a direct transition vertex between a usual
gluon and a reggeized one which can be understood as the projection of
the standard gluon field on the kinematics and polarization of the
reggeon field. Finally, for the calculations in this paper, the
induced vertex for the gluon-gluon-reggeon transition is needed. These
are graphically represented by
\begin{align}
\label{eq:1}
\includegraphics[height = 1.8cm]{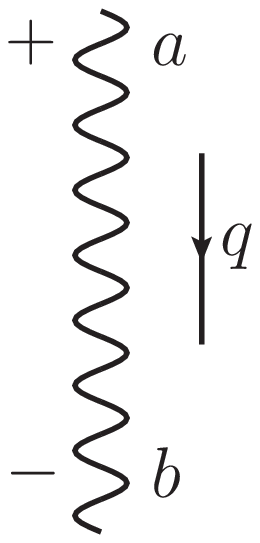} 
\qquad \qquad
 \includegraphics[height = 1.8cm]{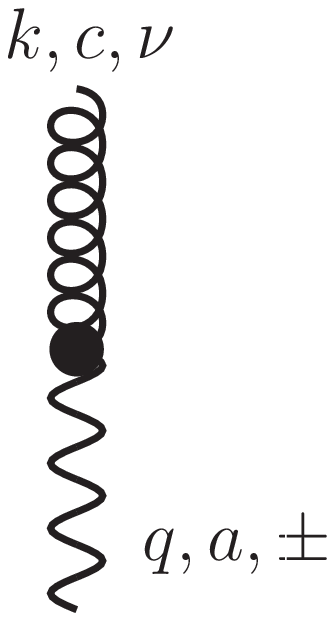}
 \qquad \qquad 
\includegraphics[height = 1.8cm]{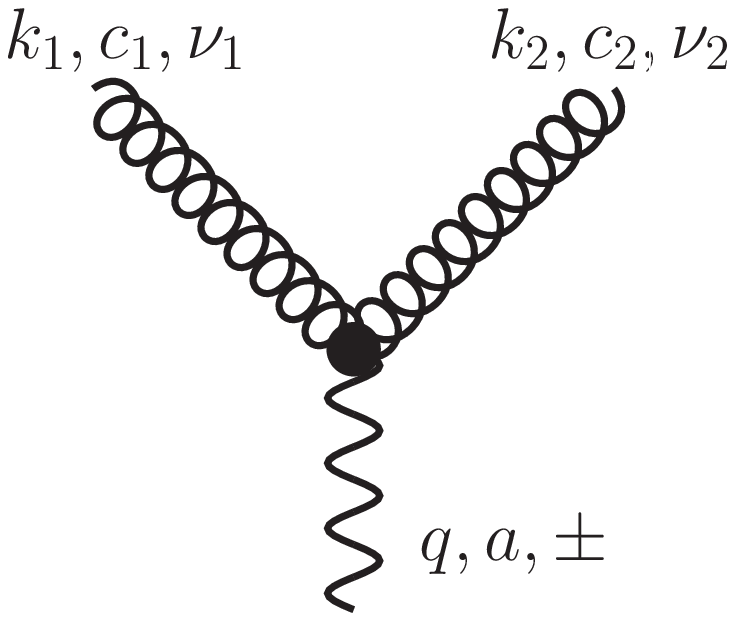}.
\end{align}
With the condition $q^\pm = 0$, the corresponding contributions are,
respectively, $\frac{i}{2 q_\perp^2} \delta^a_b$, $-i q_\perp^2
\delta^a_c (n^\pm)^\nu$ and $\frac{g}{2} f^{c_1 c_2 a} q_\perp^2
\left(\frac{1}{k_1^\pm+i \epsilon}+\frac{1}{k_1^\pm - i
    \epsilon}\right) (n^\pm)^{\nu_1} (n^\pm)^{\nu_2}$ ($k_\perp^2$
refers to Euclidean notation).  The original derivation of the
effective action does not provide a pole prescription for the eikonal
propagators of induced vertices. The above choice, which corresponds
to Cauchy principal value, has the important advantage to respect the
original Bose symmetry of the induced vertices in the effective
theory, see also the discussion in \cite{Hentschinski:2011xg}.  The
set of rules in Eq.~\eqref{eq:1} are all the pieces needed to perform the
calculation of the quark-initiated jet vertex in this work\footnote{A
  possible contribution of the three gluon - reggeon transition vertex
  to the reggeized gluon self energy discussed below in
  Sec.~\ref{sec:virtual-corrections} vanishes due to the light-like
  polarization vectors associated with this vertex. An explicit
  expression for this vertex can be found in  \cite{LevSeff,
    Hentschinski:2011xg}}.

The single tree level diagram for the four-point amplitude $p_a+p_b
\to p_1 +p_2$ with the corresponding contribution of the coupling to
the on-shell quarks reads
\begin{flushleft}
\parbox{3cm}{\includegraphics[height = 3cm]{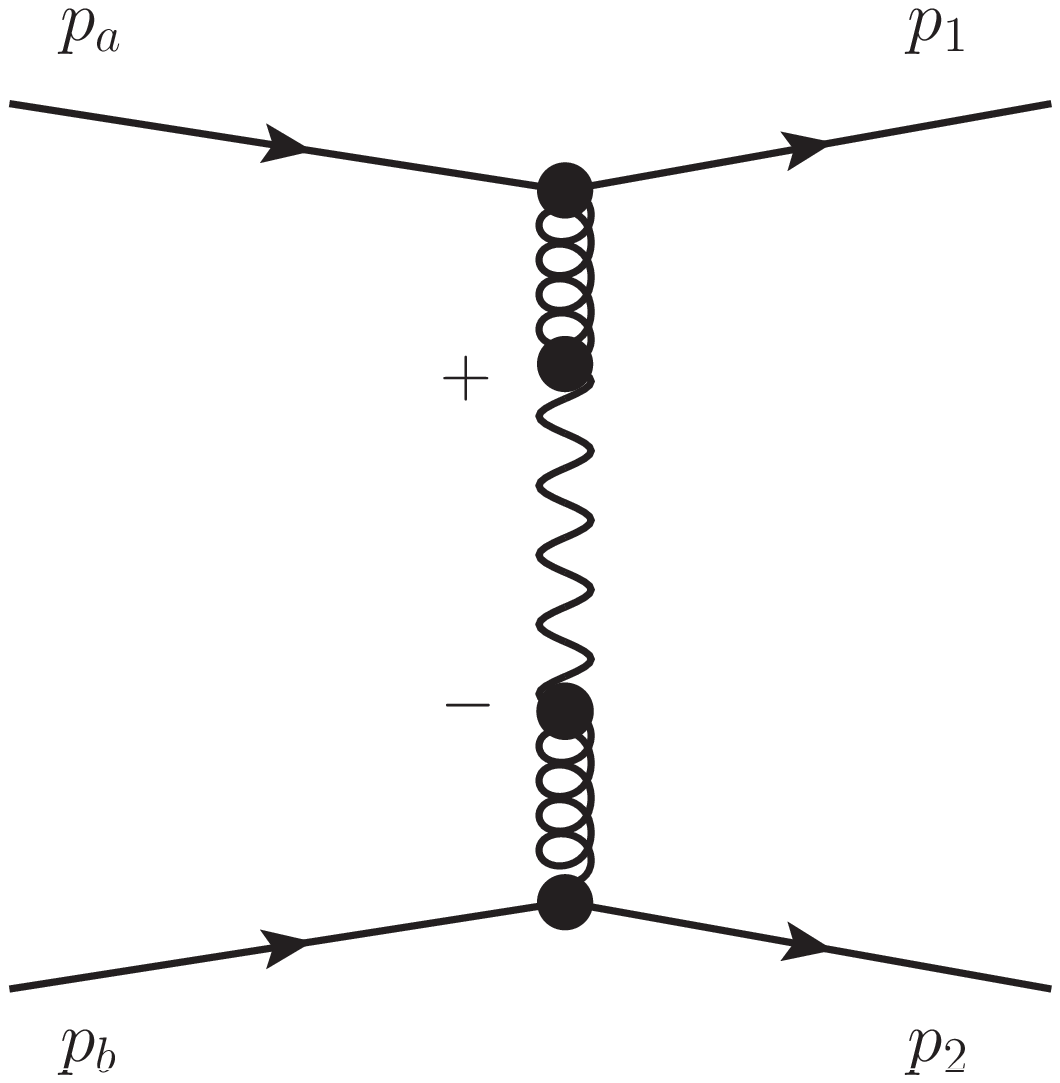}}
$~~~~ i\mathcal{M}_{qr^* \to q}^{(0)} = \hspace{-0.1cm}  
\parbox{2.5cm}{\includegraphics[width = 3cm]{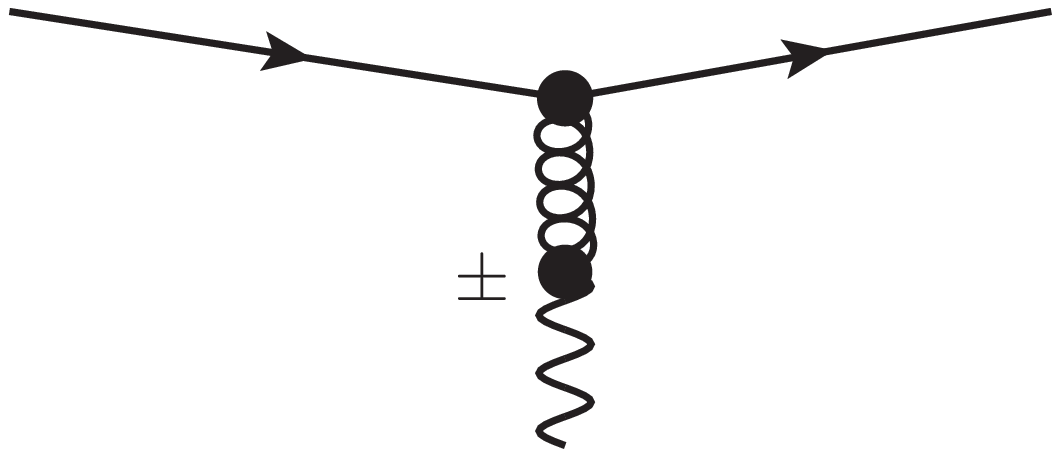}} = 
\bar{u}_{\lambda'}(p')  ig t^a \hspace{-.08cm}\not{\hspace{-.08cm}n}^\pm u(p)_\lambda =i g t^a   2 p^\pm \delta_{\lambda\lambda'}.$
\end{flushleft}
For dimension $d=4+2 \epsilon$, $C_F = \frac{N_c^2 -1}{2 N_c}$ and $\alpha_s = \frac{g^2 \mu^{2\epsilon} \Gamma(1-\epsilon)}{(4 \pi)^{1 + \epsilon}}$ the leading order quark impact factor $h^{(0)}_{a}({k_\perp}) $ at cross section level $\left(d \hat{\sigma}^{(0)}_{q_aq_b} =  h^{(0)}_{a}({k_\perp}) h^{(0)}_{b}({k_\perp}) d^{d-2} {k_\perp} \right)$ 
in $\overline{\rm MS}$ scheme is then
\begin{eqnarray}
\hspace{-1cm}&&\overline{| \mathcal{M}^{(0)}|^2}_{qr^* \to q} =  \frac{1}{4N_c (N_c^2 - 1)}\sum_{\lambda\lambda'}| \mathcal{M}^{(0)}|^2_{qr^* \to q}  =  \frac{ 4 g^2  C_F}{{N_c^2 -1}}   p_a^{+2}, \\
\hspace{-1cm}&&h_a^{(0)}(k_\perp) = \frac{\sqrt{N_c^2 - 1}}{(2 p_a^+)^2} \int \frac{d k^-}{(2 \pi)^{2 + \epsilon}} \int   
\frac{d \Phi^{(1)} }{k_\perp^2} \overline{| \mathcal{M}^{(0)}|^2}_{qr^* \to q} =
\frac{C_F}{\sqrt{N_c^2 -1}} \frac{2^{1+\epsilon}}{\mu^{2\epsilon} \Gamma(1-\epsilon)} \frac{1}{k_\perp^2},
\end{eqnarray}
which is in agreement with~\cite{Ciafaloni:1998hu,Bartels:2001ge}.

\section{Virtual Corrections}
\label{sec:virtual-corrections}
When calculating quantum corrections to the tree level result discussed above new divergencies in longitudinal 
components appear. In the present work these are regularized by going away from the light cone using a 
parameter $\rho$ which is considered in the limit $\rho \to \infty$ and can be  
interpreted as a logarithm of the center-of-mass energy. In particular, the 
Sudakov projections take place on the vectors $n_a = e^{-\rho} n^+ + n^-$ and $n_b = n^+ + e^{- \rho} n^-$. To 
study the virtual corrections it is needed to obtain the one-loop self energy corrections to the reggeon 
propagator. Diagrammatically these are
\begin{figure}[htb]
  \centering
  \parbox{2cm}{\vspace{0.1cm} \includegraphics[height = 2cm]{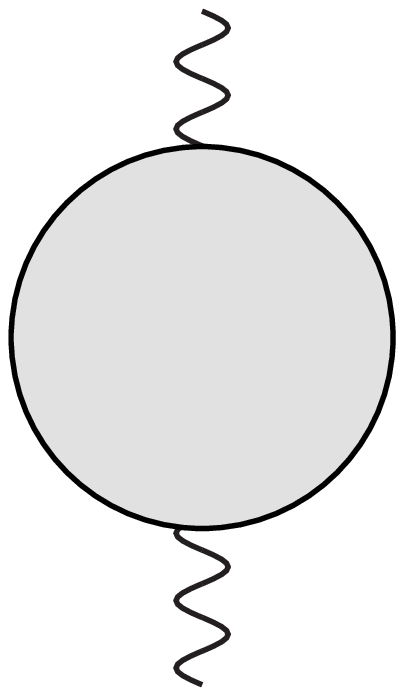}} \hspace{-0.8cm}= 
  $\left[
  \parbox{1.5cm}{\vspace{0.1cm} \includegraphics[height = 2cm]{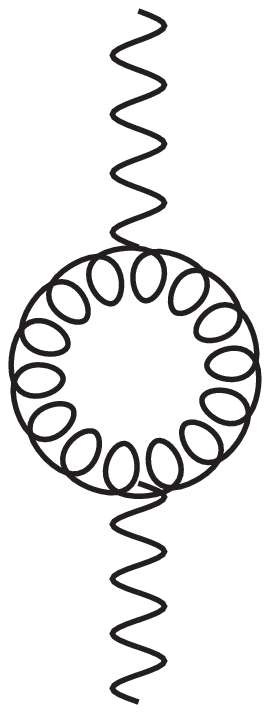}} \hspace{-0.6cm}\right]_1$
  + \hspace{0.1cm} $\left[
  \parbox{1.5cm}{\vspace{0.1cm} \includegraphics[height = 2cm]{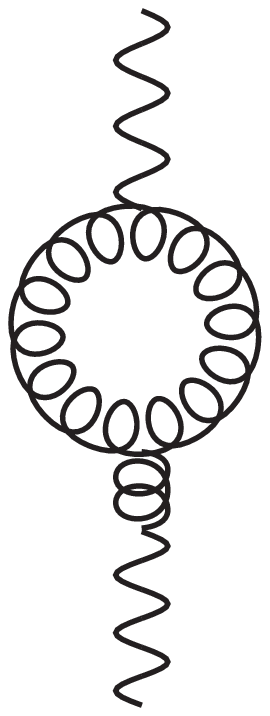}} 
 \hspace{-0.6cm} +
  \parbox{1.5cm}{\vspace{0.1cm} \includegraphics[height = 2cm]{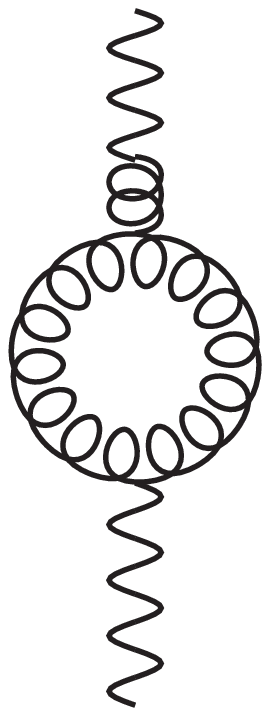}} \hspace{-0.6cm}\right]_2$
 + \hspace{0.1cm} $\left[
  \parbox{1.5cm}{\vspace{0.1cm} \includegraphics[height = 2cm]{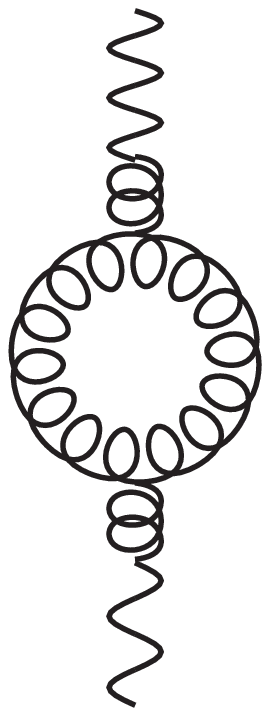}} 
 \hspace{-0.6cm}+
  \parbox{1.5cm}{\vspace{0.1cm} \includegraphics[height = 2cm]{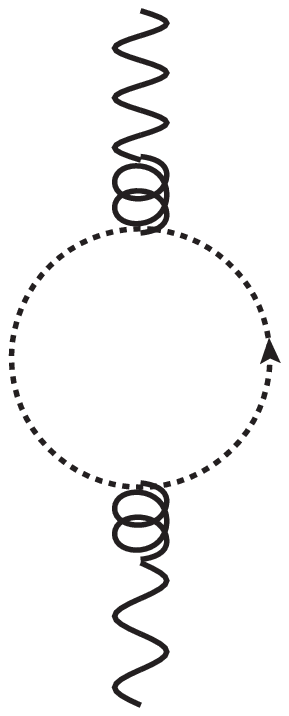}}\hspace{-0.6cm}\right]_3$
\hspace{0.1cm} + $\left[
  \parbox{1.5cm}{\vspace{0.1cm} \includegraphics[height = 2cm]{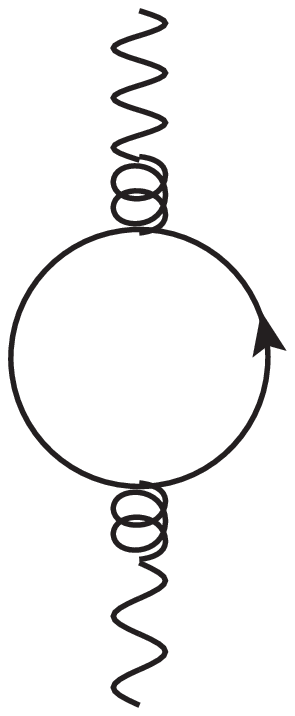}} \hspace{-0.6cm}\right]_4$\\ \hspace{.98\textwidth}.
\end{figure}\\
In $[\dots]_3$ a ghost loop has been included. Keeping the leading $\rho$ terms, using $\beta_0=(11 N_c -2 n_f)/3$  
and $\lambda \equiv \frac{ (-2 i k_\perp^2) g^2 N_c }{(4\pi)^{2 + \epsilon}}  \left(\frac{k_\perp^2}{\mu^2} \right)^\epsilon  \frac{\Gamma(1 - \epsilon)\Gamma(1 + \epsilon)^2}{\Gamma(1 + 2 \epsilon)}$, 
each contribution and the full result are
\begin{eqnarray}
\parbox{2.5cm}{\includegraphics[width = 1cm]{traj.eps}} \hspace{-1.8cm} &=& \hspace{-.3cm} 
\lambda  \bigg\{ \bigg[ \frac{ i\pi - 2 \rho }{  \epsilon}  \bigg]_1       
+   \frac{1}{(1 + 2 \epsilon)\epsilon} \bigg(\bigg[-2 \bigg]_2
 - \bigg[ \frac{5 + 3\epsilon}{3 + 2 \epsilon} \bigg]_3
+ \frac{n_f}{N_c}  \bigg[ \frac{2 + 2\epsilon}{3 + 2\epsilon}\bigg]_4 \bigg) \bigg\}  \nonumber\\
&=& \hspace{-.3cm} \frac{\alpha_sN_c (-2i k_\perp^2)}{  2\pi    }  \left(\frac{k_\perp^2}{\mu^2} \right)^\epsilon   
\bigg( 
\frac{i\pi - 2 \rho  }{2 \epsilon} - \frac{\beta_0}{ 2 N_c  \epsilon} + \frac{67 }{18 } - \frac{10 n_f }{18 N_c} \bigg) + \mathcal{O}(\epsilon).
\end{eqnarray}
From the effective action the one-loop corrections to the $ i \mathcal{M}^{(1)}_{qr^* \to q}$ quark-quark-reggeon vertex are
\begin{flushleft}
\parbox{2cm}{\includegraphics[width = 2cm]{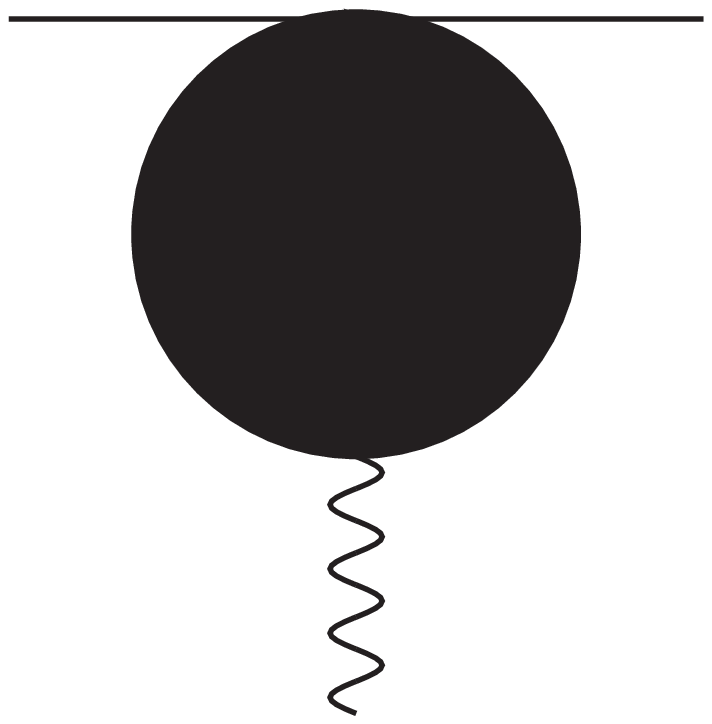}}  =
  $\left[\parbox{2cm}{\includegraphics[width = 2cm]{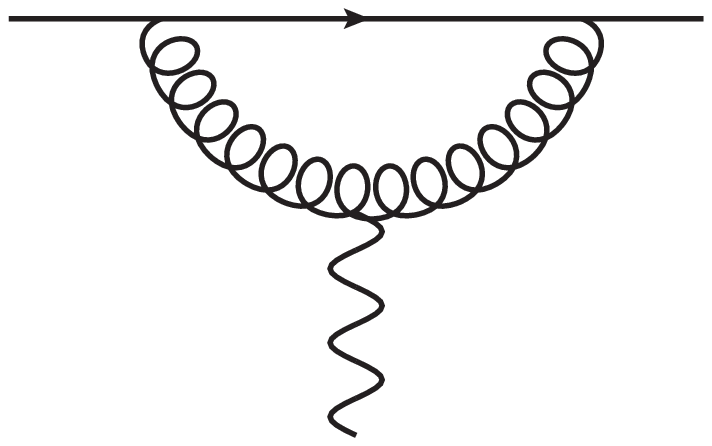}} \right]_1$
  + 
  $\left[\parbox{2cm}{\includegraphics[width = 2cm]{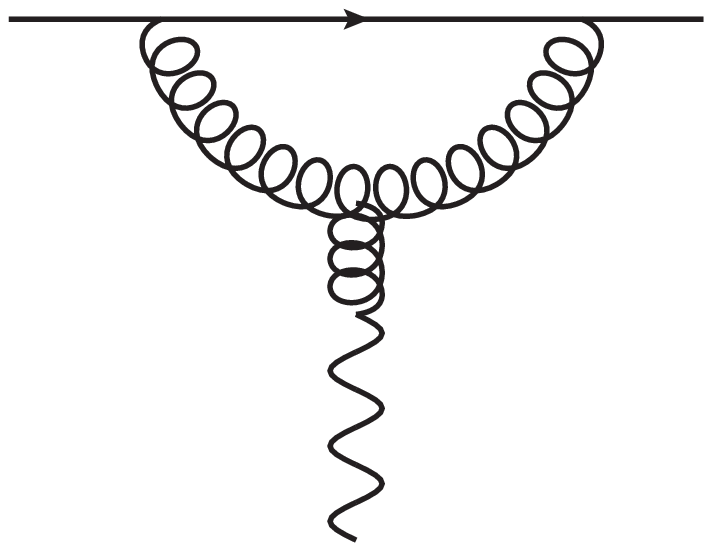}} \right]_2$
  +
  $\left[\parbox{2cm}{\includegraphics[width = 2cm]{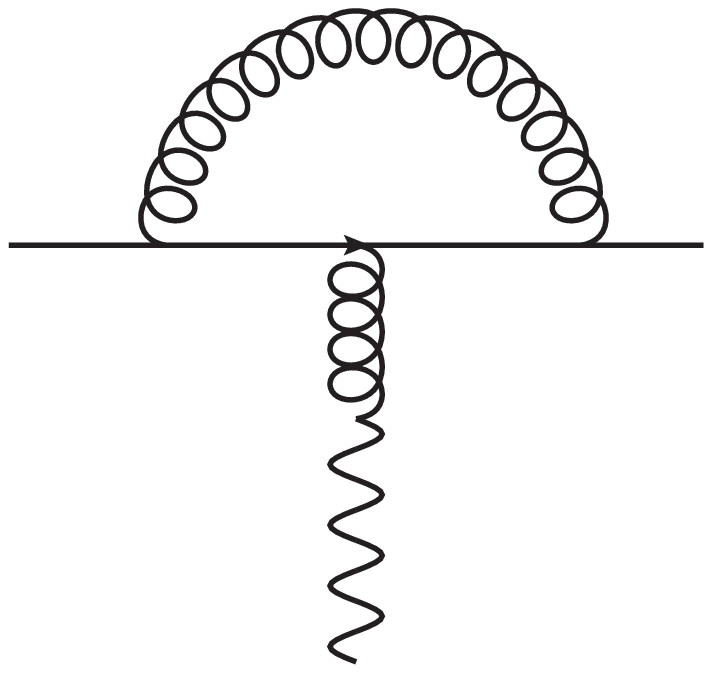}} \right]_3$\\
\hspace{2cm} + 
$\left[\parbox{2cm}{\includegraphics[width = 2cm]{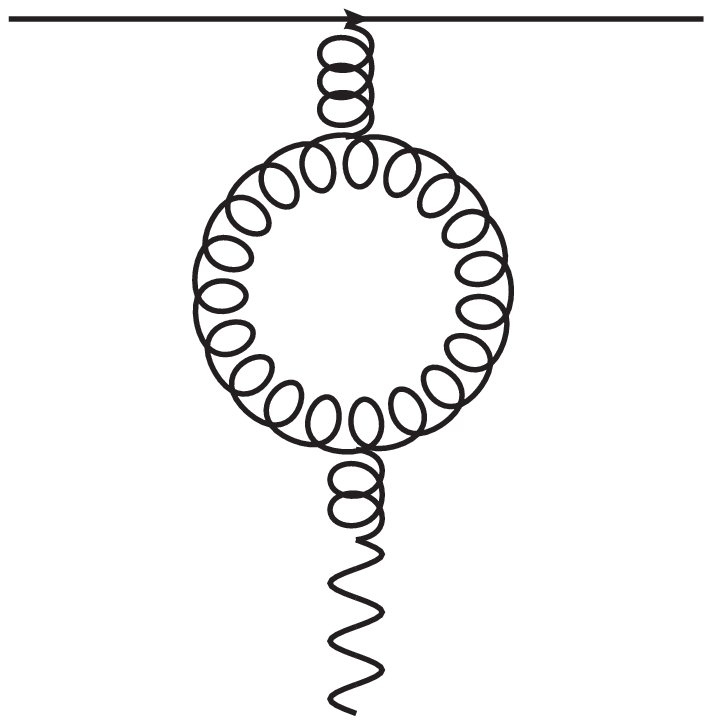}} 
+
\parbox{2cm}{\includegraphics[width = 2cm]{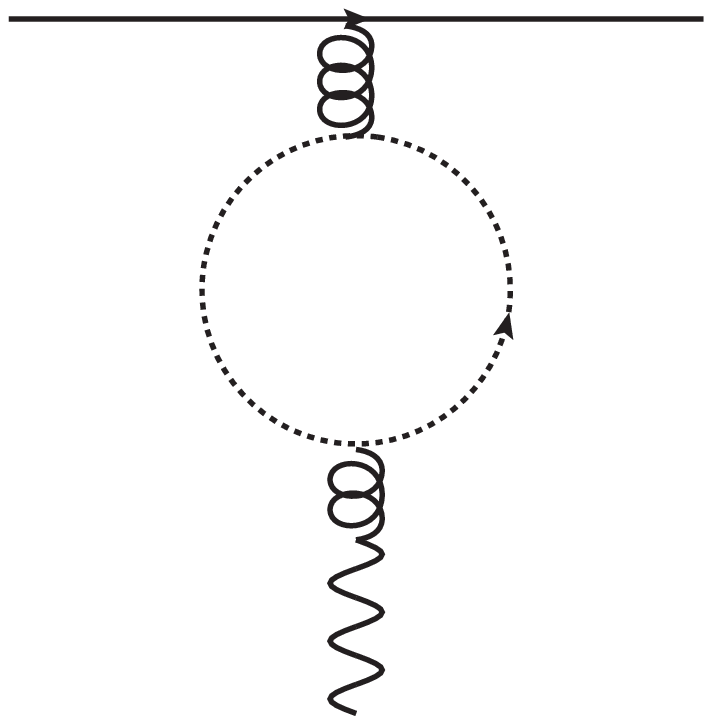}} \right]_4$
+
$\left[\parbox{2cm}{\includegraphics[width = 2cm]{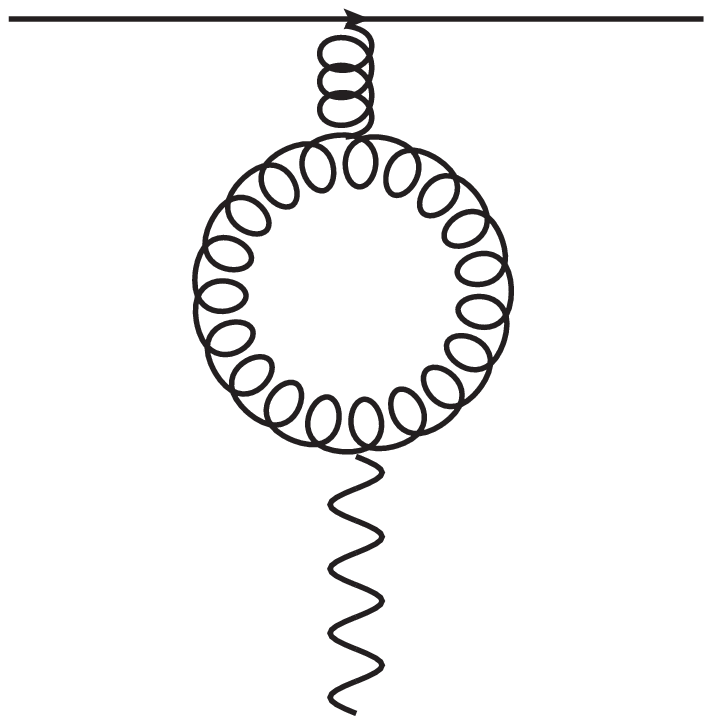}} \right]_5$
+
$\left[\parbox{2cm}{\includegraphics[width = 2cm]{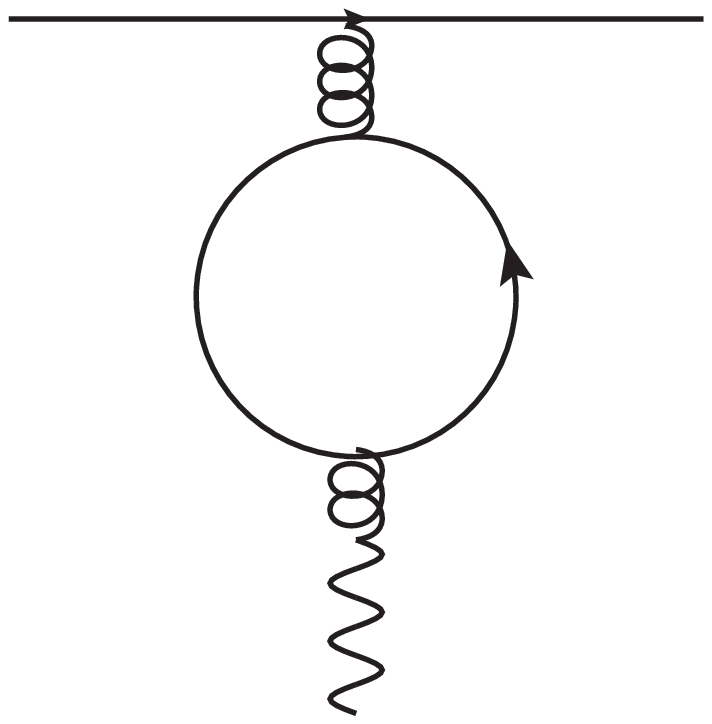}} \right]_6$
\begin{eqnarray}
\hspace{-0.4cm}&=& \frac{i \mathcal{M}^{(0)}_{qr^* \to q}}{(-2i k_\perp^2)} \frac{\lambda}{\epsilon} \bigg\{
\bigg[\frac{1}{\epsilon}  - 2 \ln \frac{p_a^+}{\sqrt{k_\perp^2}} -2 \rho + i\pi - \psi(1-\epsilon)  - \psi(1)    + 2 \psi(\epsilon) \bigg]_1 \nonumber\\
&&\hspace{-0.2cm}+ \frac{1}{(1 + 2\epsilon)} 
\bigg( \bigg[ \frac{1}{2}  \bigg]_2
+ \frac{1}{N_c^2}  \bigg[\frac{2 + \epsilon + 2\epsilon^2}{2\epsilon}  \bigg]_3
- \bigg[ \frac{3\epsilon + 5}{3 + 2\epsilon} \bigg]_4
- \bigg[ 1 \bigg]_5 
+2  \frac{n_f}{N_c}  \bigg[ \frac{1+\epsilon }{3 + 2\epsilon} \bigg]_6
 \bigg]
 \bigg\} \nonumber\\
&&\hspace{-0.6cm}= \frac{i \mathcal{M}^{(0)}_{qr^* \to q}}{(-i2 k_\perp^2)}  \frac{\lambda}{\epsilon}
\bigg\{
-  
\left( \ln \frac{-p_a^+}{\sqrt{k_\perp^2}} +   \ln \frac{p_a^+}{\sqrt{k_\perp^2}} + {\rho} \right) 
 - 
\frac{1}{(1 + 2\epsilon)}
 \bigg[  
-\frac{1}{N^2_c} 
\left(  \frac{1}{\epsilon} + \frac{1 + 2\epsilon}{2} \right)  \nonumber\\
&&\hspace{-0.2cm}+  \frac{11 + 7 \epsilon}{3 + 2\epsilon} - 2 \frac{n_f}{N_c} \frac{1 + \epsilon}{3 + 2\epsilon}  
-  \frac{2 + 7 \epsilon}{2 \epsilon} +  (1 + 2\epsilon) 
\bigg(
\psi(1-\epsilon) - 2 \psi(\epsilon) + \psi(1)
\bigg)
 \bigg]  
\bigg\},
\end{eqnarray}
\end{flushleft}
where $ \psi(z) = {\Gamma'(z)}/{\Gamma(z)}$.  The original
presentation in \cite{LevSeff} defines the effective action to be
local in rapidity space, with all non-local interactions mediated
through reggeon exchange alone.  In the case of tree-level amplitudes
in the QMRK locality in rapidity is naturally imposed through explicit
cut-offs which force the final state particles into rapidity clusters
strongly ordered in rapidity with respect to each other.  As an
alternative to such explicit cut-offs it is possible to subtract from
all `local' matrix elements (which do not contain reggeon exchange)
the corresponding `non-local' contribution with reggeon exchange. The
resulting subtracted matrix element then vanishes in the limit of large
internal internal rapidity separations up to sub-leading
corrections. They are therefore already local in rapidity space and a
cut-off is no longer needed. In addition, the subtraction avoids, by construction, possible 
over-counting due to iteration of lower order contributions and allows
to verify the equivalence of the QCD and effective amplitudes at each order of
perturbation theory.  In the present case we subtract from the above
result the non-local contributions stemming from the one-loop
corrections to the reggeon propagator combined with the tree-level
quark-reggeon coupling:
\begin{eqnarray}
\hspace{-1cm}\parbox{2cm}{\includegraphics[width = 2cm]{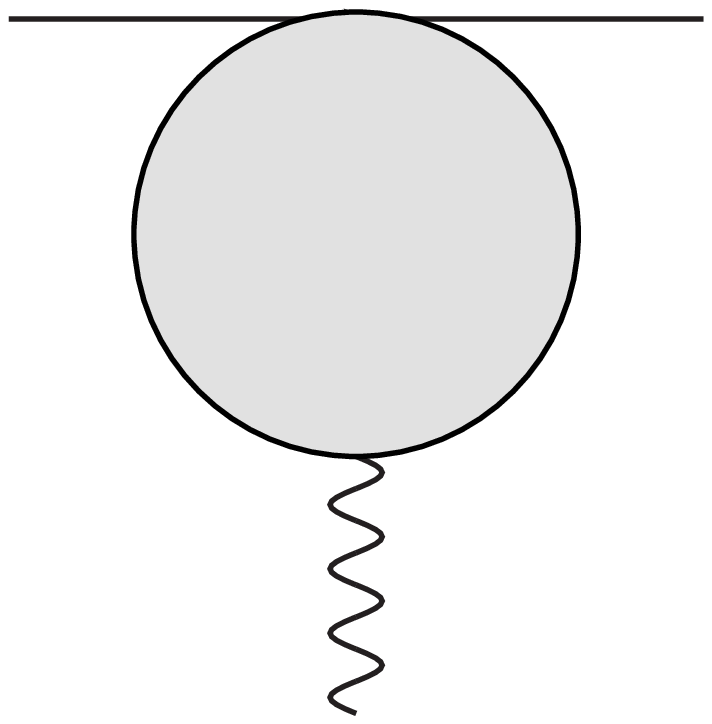}} &=&  
\parbox{2cm}{\includegraphics[width = 2cm]{impaamp.eps}} 
- \parbox{2cm}{\includegraphics[width = 2cm]{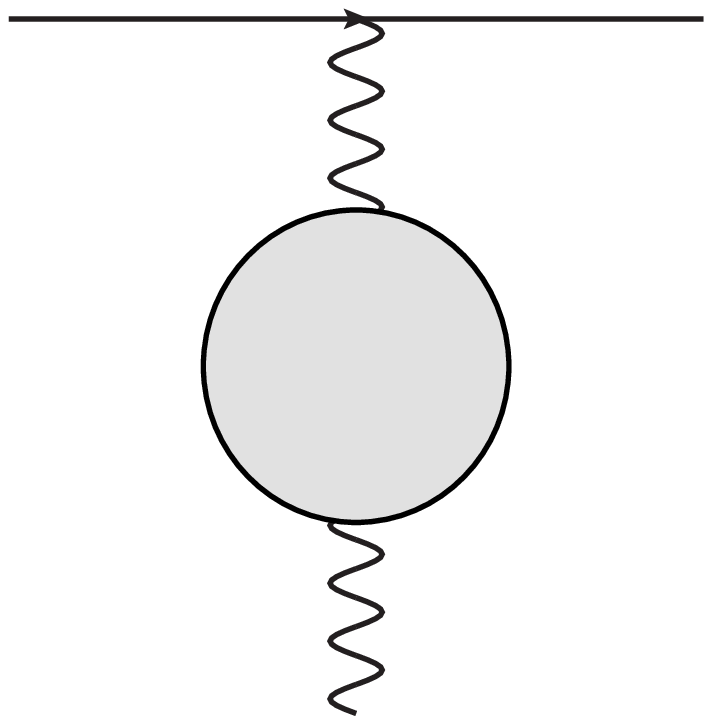}} \nonumber\\
&& \hspace{-2cm} = i \mathcal{M}^{(0)}_{qr^* \to q}   \frac{g^2  }{(4\pi)^{2 + \epsilon}}  
\left(\frac{ k_\perp^2}{\mu^2} \right)^\epsilon  \frac{\Gamma(1-\epsilon) \Gamma^2 (1+\epsilon)}{\Gamma(1+2 \epsilon)}  \left\{
 \frac{-2 N_c}{\epsilon}  
\left(  \ln \frac{p_a^+}{\sqrt{k_\perp^2}} - \frac{\rho}{2} \right)  \right. \nonumber\\
&& \hspace{-1cm}\left. +
\frac{ N_c (2 + 7 \epsilon)}{2\epsilon^2(1 + 2\epsilon)} + 
 \frac{1}{N_c} 
\left(  \frac{1}{\epsilon^2 (1 + 2 \epsilon)} + \frac{1}{2 \epsilon} \right)  - N_c\frac{1}{\epsilon}  \bigg(\psi(1-\epsilon) - 2 \psi(\epsilon) + \psi(1)\bigg)\right\}.
\end{eqnarray}
The four-point elastic amplitude is the sum of two contributions as the one calculated above:
\begin{eqnarray}   
i\mathcal{M}_{q_a q_b \to q_1 q_2}^{(1)} =  \parbox{2cm}{\includegraphics[width = 2cm]{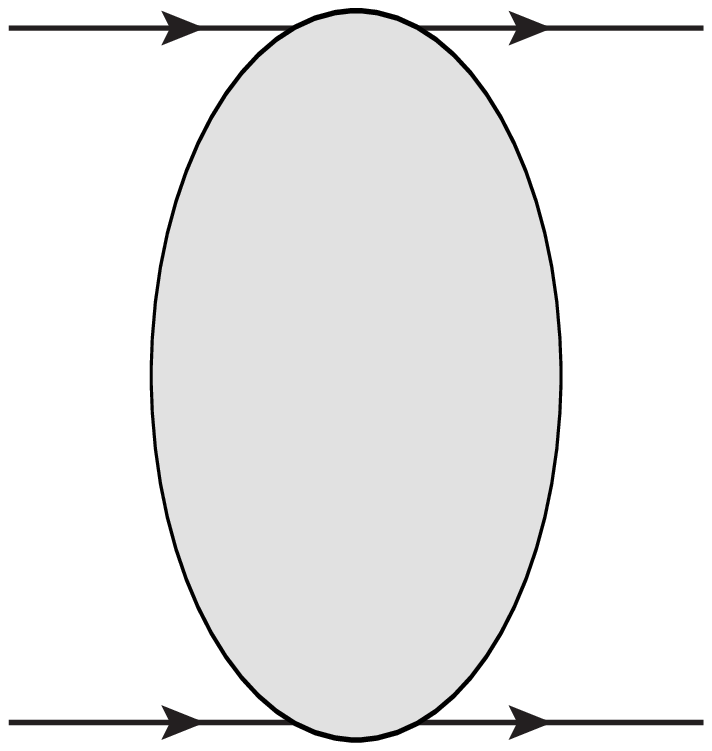}} &=  \parbox{2cm}{\includegraphics[width = 2cm]{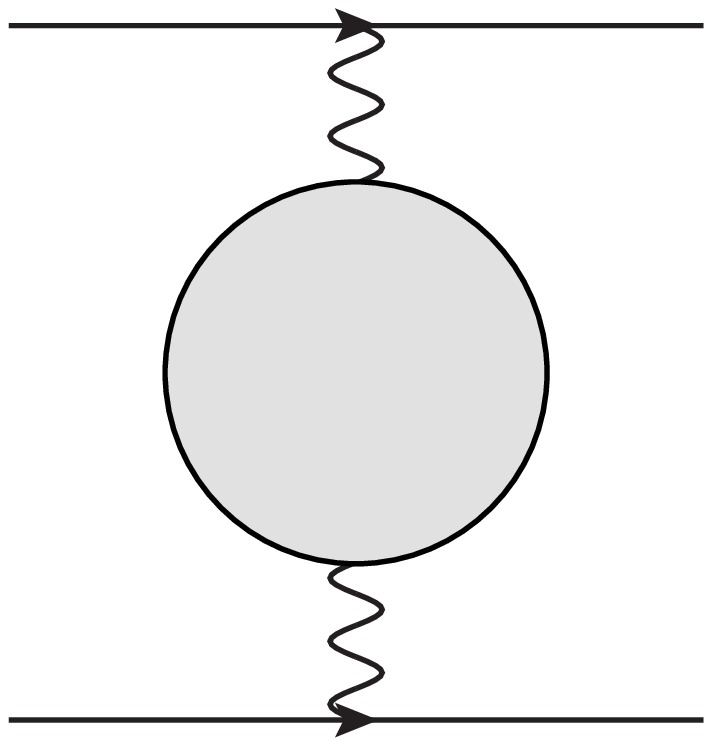}} +  \parbox{2cm}{\includegraphics[width = 2cm]{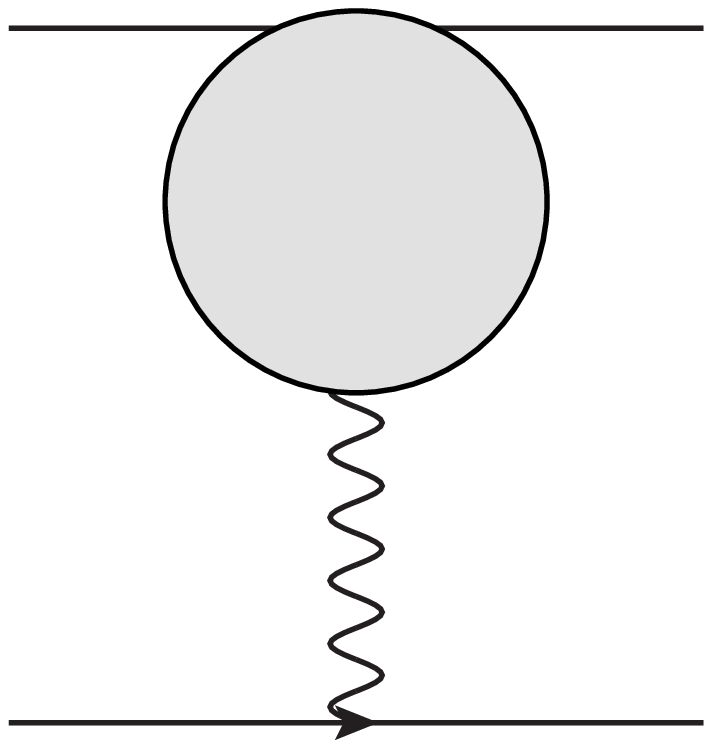}} +  \parbox{2cm}{\includegraphics[width = 2cm]{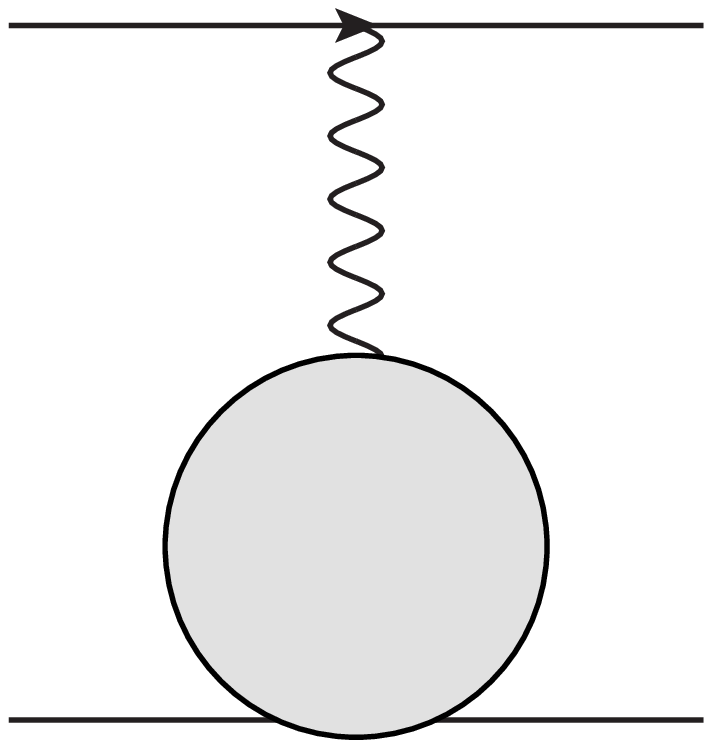}}, \nonumber
\end{eqnarray}
where the $\rho$ dependence cancels while the terms with $\ln p_a^+/\sqrt{k_\perp^2}$ and  
$\ln p_b^-/\sqrt{k_\perp^2}$ generate the LL contribution of the form 
($\omega(- k_\perp^2) = - \frac{\alpha_s N_c}{2 \pi} \left(\frac{1}{\epsilon}+\ln{\frac{k_\perp^2}{\mu^2}}\right)$ is the LL gluon Regge trajectory)
\begin{eqnarray}
i\mathcal{M}_{q_a q_b \to q_1 q_2}^{(1)} &=&  i\mathcal{M}_{q_a q_b \to q_1 q_2}^{(0)}  
\bigg(\frac{1}{2} \left( \ln \frac{s}{ k_\perp^2}+  \ln \frac{-s}{ k_\perp^2} \right) \omega(- k_\perp^2)  
+ \Gamma_a^{(1)} ( k_\perp^2) + \Gamma_b^{(1)} ( k_\perp^2) \bigg).
\end{eqnarray}
This piece is associated to the reggeon exchange and is removed from the quark-quark-reggeon vertex corresponding to the coupling to the external fields, which hence reads
\begin{align}
 \Gamma^{(1)}_{a,b} ( k_\perp^2) &=\!
\frac{\alpha_s}{2} \left(\frac{k_\perp^2}{\mu^2} \right)^\epsilon \!
\left[\frac{N_c}{\pi} \left( \frac{85}{36} + \frac{\pi^2}{4} \right)-\frac{\beta_0}{4 \pi \epsilon} - \frac{5}{18} \frac{n_f}{\pi}
-\frac{C_F}{\pi} \left(  \frac{1}{\epsilon^2}  - \frac{3}{2\epsilon} + 4 - \frac{\pi^2}{6}\right) \right].
\end{align}
This result is in perfect agreement with the more standard calculations, not using the high energy effective action,  performed in~\cite{Fadin:1993qb} and confirmed in~\cite{DelDuca:1998kx}. 

\section{Real Corrections}

The real corrections to the Born level process can be organized into three contributions to the five-point 
amplitude with central and quasi-elastic gluon production, {\it i.e.},
\begin{figure}[htb]
  \centering
  \parbox{2.5cm}{\center \includegraphics[width = 2cm]{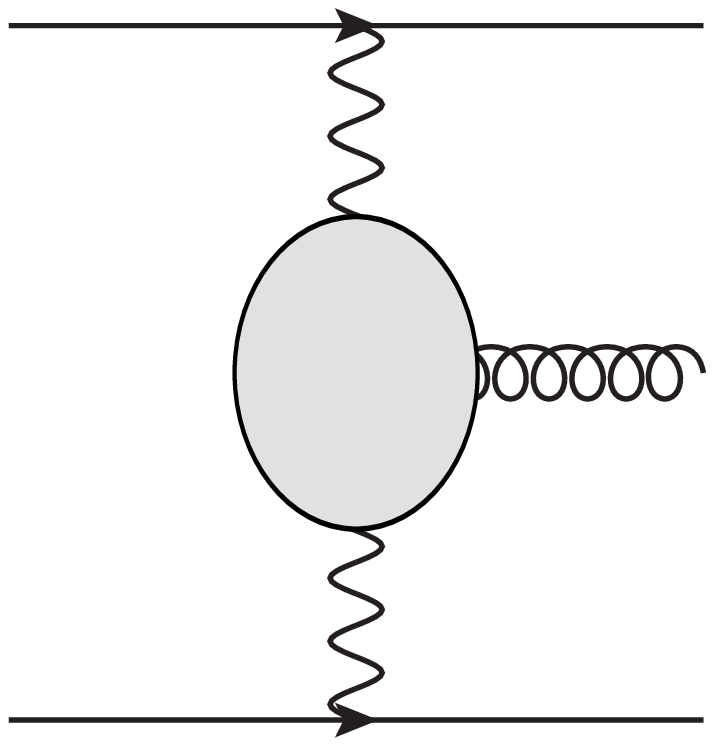}}
 \parbox{2.5cm}{\center \includegraphics[width = 2cm]{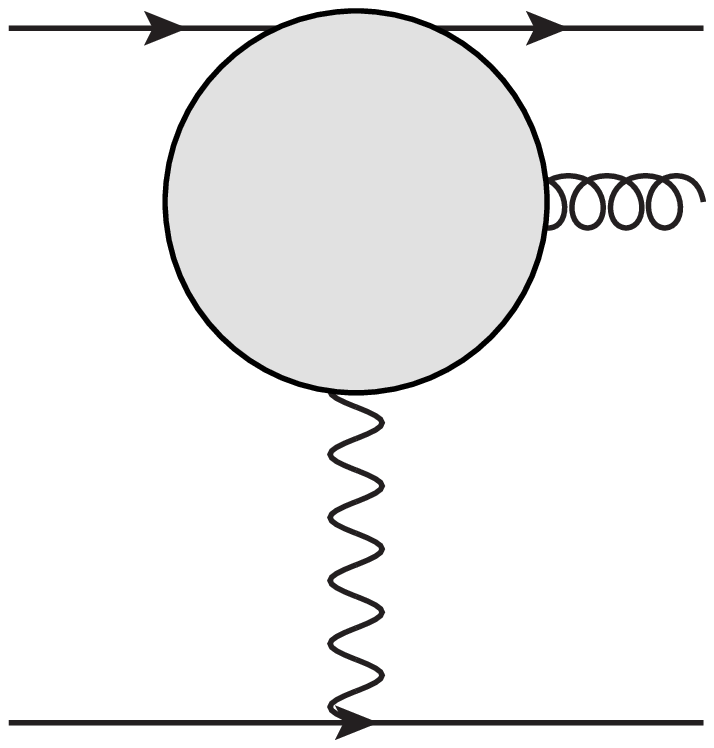}}
 \parbox{2.5cm}{\center \includegraphics[width = 2cm]{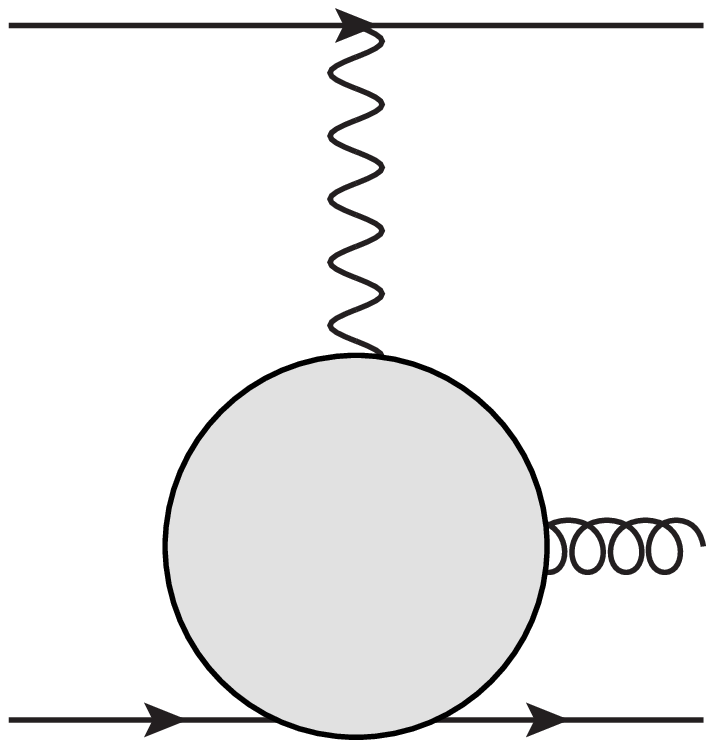}}
\end{figure}\\
To regularize the divergencies appearing in the integration over the longitudinal phase space of the gluon an 
explicit cut-off in rapidity will be used for the intermediate steps in the calculation.  

The central production amplitude 
yields the unintegrated real part of the forward leading order BFKL kernel and is obtained from the sum of 
the following three effective diagrams:
\begin{eqnarray}
   \parbox{2cm}{\center \includegraphics[width=1.2cm]{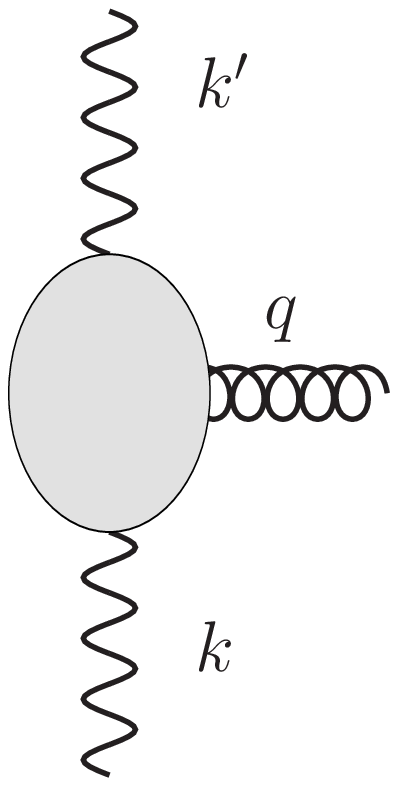}} 
& =&
   \parbox{2cm}{\center \includegraphics[width=1.2cm]{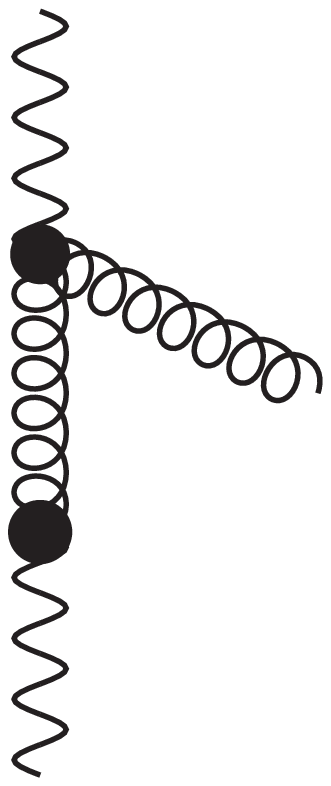}} +
   \parbox{2cm}{\center \includegraphics[width=1.2cm]{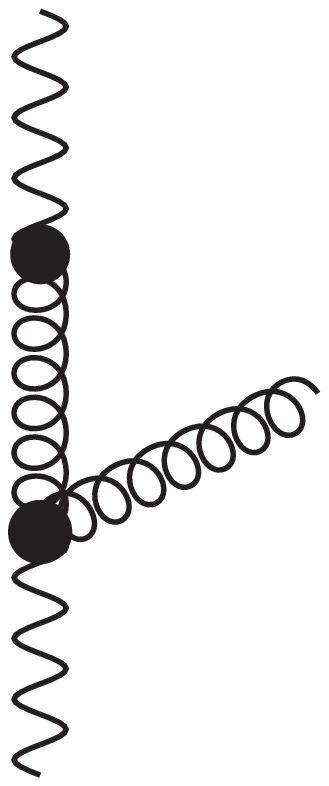}} +
   \parbox{2cm}{\center \includegraphics[width=1.2cm]{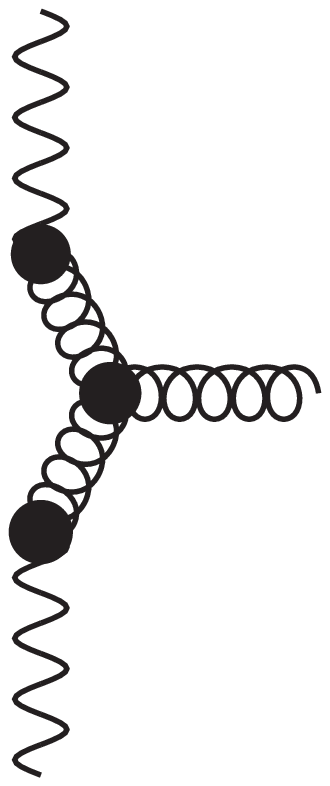}}. \nonumber
\end{eqnarray}
The Sudakov decomposition of the external momenta is $k' = q^+ n^+ / 2
+ q_\perp -k_\perp$, $k = k^- n^+ / 2 + k_\perp$ and $q = q^+ n^+ /2 +
k^- n^+ / 2 + q_\perp$, with $k^- = q_\perp^2 / q^+$, where ${k'}^- =
0 = k^+$ is a direct consequence of Eq.~\eqref{eq:constraint}.  The
squared amplitude, averaged over color of the incoming reggeons and
summed over final state color and helicities reads
\begin{eqnarray} 
  \overline{|\mathcal{M}|^2}_{r^*r^* \to g} &=& \frac{16g^2 N_c}{N_c^2 - 1} 
         \frac{( q_\perp -  k_\perp )^2 k_\perp^2 }{ q_\perp^2}.
\end{eqnarray}
It leads to the following production vertex
\begin{eqnarray}
  V(k_\perp, k'_\perp) &=& 
  \frac{N_c^2 - 1}{8 (2 \pi)^{3+2 \epsilon} k_\perp^{' 2}  k_\perp^2}
  \overline{|\mathcal{M}|^2}_{r^*r^* \to g} ~=~   \frac{\alpha_s N_c}{\pi_\epsilon \pi({k_\perp} +  k'_\perp)^2} , 
\end{eqnarray}
with $\pi_\epsilon \equiv \pi^{1 + \epsilon} \Gamma(1 - \epsilon) \mu^{2 \epsilon}$ and momentum conservation and on-shellness implied. Finally, the central production 
contribution to the exclusive differential cross section is
\begin{eqnarray}
 d \hat{\sigma}^{(c)}_{ab} &=& h^{(0)}_a({k'_\perp}) h^{(0)}_b({k_\perp})    
 \mathcal{V}({q_\perp}; {k_\perp}, {k'_\perp})  d^{2+2\epsilon} {k'_\perp} d^{2+2\epsilon} {k_\perp} d \eta,
  \end{eqnarray}
with ${\cal V}({q_\perp}; {k_\perp}, \eta_a, \eta_b) \equiv 
V({q_\perp}; {k_\perp}, {k'_\perp})  \theta(\eta_a - \eta)\theta(\eta - \eta_b)$ being the regularized production vertex.

The quasi-elastic contribution $q(p_a) r^*(k) \to g(q)q(p)$ is the sum of the effective diagrams
\begin{eqnarray}
 \parbox{4cm}{\center \includegraphics[width=3.5cm]{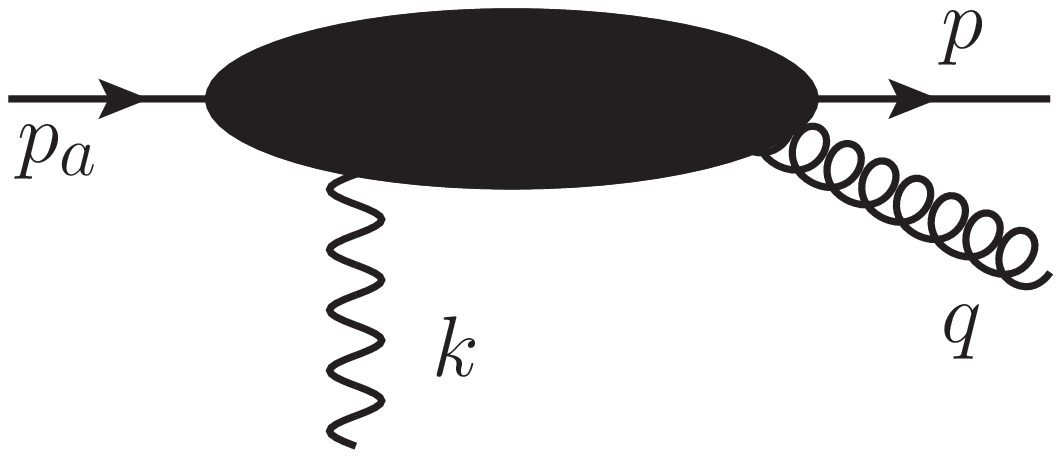}} \hspace{-0.5cm}&=&
\hspace{-0.6cm}  \parbox{4cm}{\center \includegraphics[width=3.5cm]{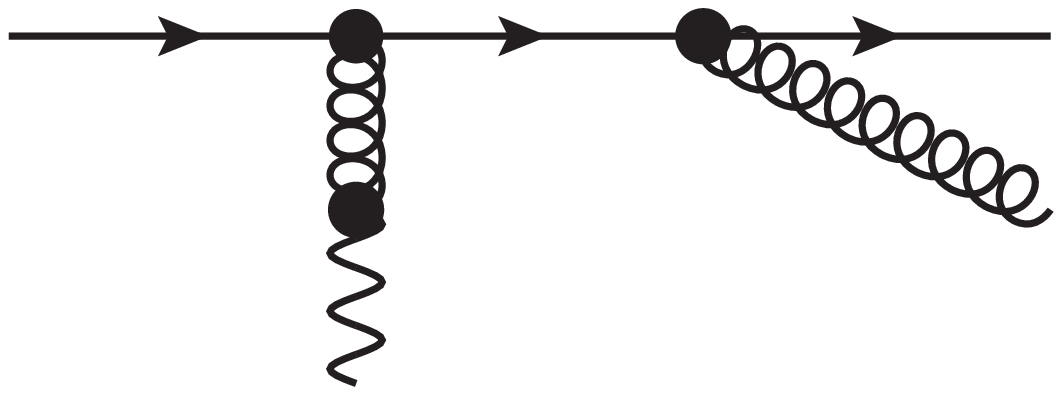}} \hspace{-0.3cm}+
\hspace{-0.3cm}   \parbox{4cm}{\center \includegraphics[width=3.5cm]{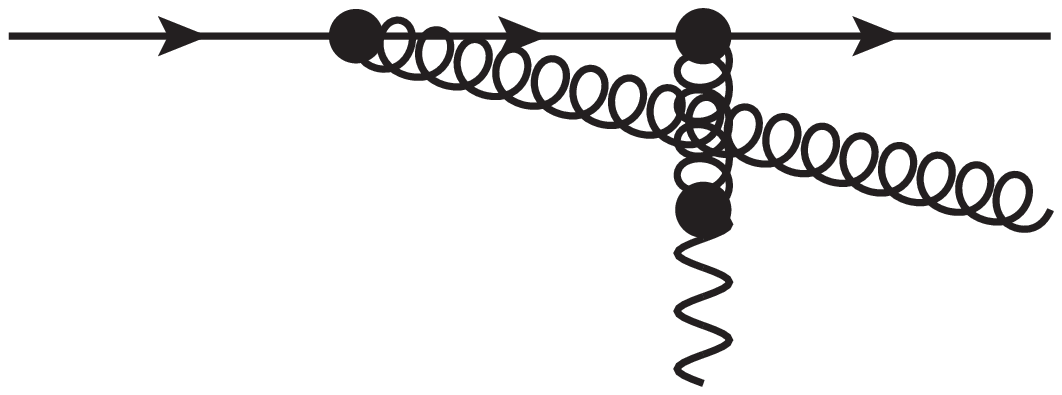}} \hspace{-0.3cm} \nonumber \\
&+&
\hspace{-0.3cm} \parbox{4cm}{\center \includegraphics[width=3.5cm]{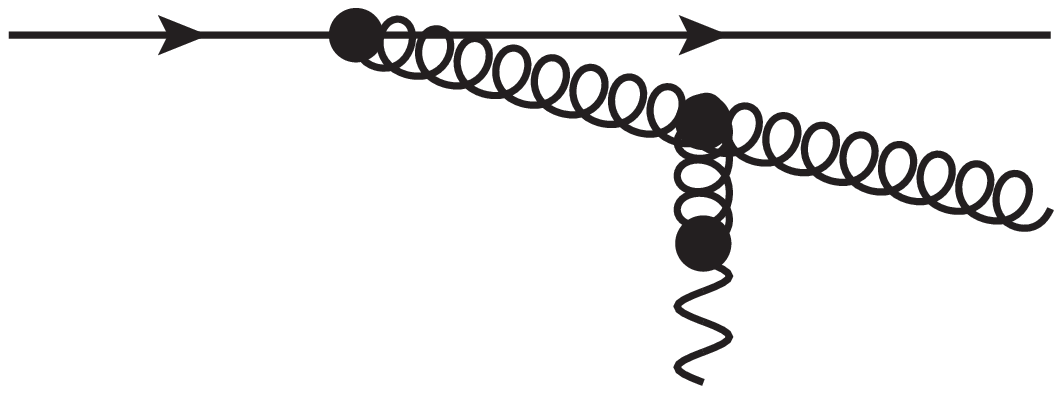}} \hspace{-0.3cm} +
\hspace{-0.3cm} \parbox{4cm}{\center \includegraphics[width=3.5cm]{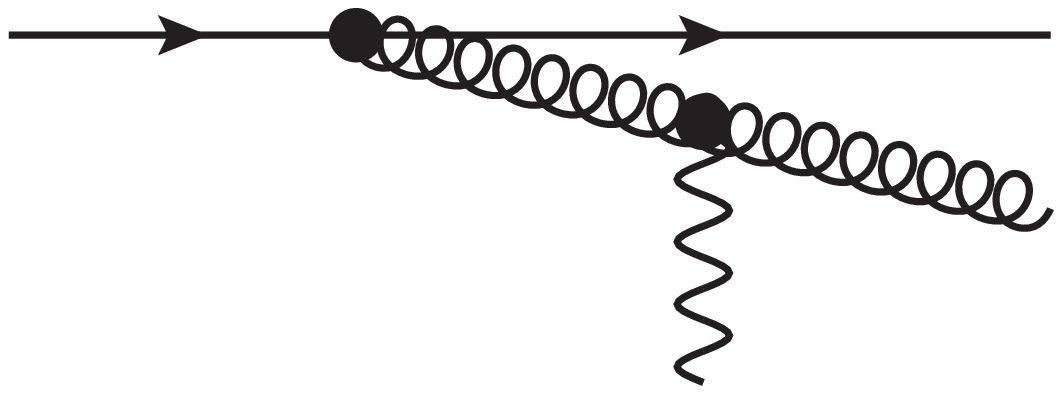}} \nonumber
\end{eqnarray}
The Sudakov decomposition $p = (p_a^+ - q^+) n^- /2 + p^- n^+ /2 + k_\perp - q_\perp$, 
$q = q^+ n^- /2 + (k^- - p^-) n^+ /2 + q_\perp$, with $p^- = (k_\perp - q_\perp)^2 / ((1-z) p_a^+)$, 
$k^- = (\Delta_\perp^2+z(1-z)k_\perp^2)/((1-z)z p^+_a)$, $z= q^+ / p^+_a$ and 
$\Delta_\perp \equiv q_\perp - z k_\perp$, 
is used to obtain the squared amplitude
\begin{eqnarray}
\overline{|\mathcal{M}|^2}_{r^*q \to qg}& =& 
\frac{g^4 8 p_a^{+2}}{{N_c^2 -1}} \frac{\mathcal{P}_{gq}(z,\epsilon)}{\Delta_\perp^2 q_\perp^2}  \frac{(1-z)z k_\perp^2}{{k^{'2}_\perp}} 
\theta\left(z - e^{-\eta_b}\frac{\sqrt{q_\perp^2}}{p_a^+} \right)
 \nonumber \\ && \qquad \qquad \qquad \qquad \cdot
\left[C_F z^2 {k^{'2}_\perp} + N_c (1-z) {\Delta_\perp} \cdot {q_\perp} \right]
,
\end{eqnarray}
where $\mathcal{P}_{gq}(z,\epsilon) = C_F \frac{1 + (1-z)^2 + \epsilon z^2}{z}$ is the real part of the $q \to g$ splitting function and  $k_\perp' = q_\perp - k_\perp$.   $\eta_b$ is a lower cut-off in rapidity. To wrap up, the real corrections to the impact factor 
are then 
\begin{eqnarray}
h^{(1)}({k_\perp}) dz d^{2+2\epsilon}{q_\perp}  &=&  \frac{\sqrt{N_c^2 - 1}}{(2 p_a^+)^2} 
\int \frac{d k^-}{(2 \pi)^{2 + \epsilon}}  d \Phi^{(2)} | \mathcal{M}|^2_{qg^* \to qg} \frac{1}{k_\perp^2},
\end{eqnarray}
with the two-particle phase space being
\begin{eqnarray}
d\Phi^{(2)} &=&  \frac{1}{2p_a^+ (2\pi)^{2 + 2\epsilon}} dz d^{2+2\epsilon}{q_\perp} 
\frac{1}{(1-z)z} \delta\left(k^- - \frac{{\Delta_\perp}^2 + z (1-z) k_\perp^2}{(1-z)p_a^+}\right).
\end{eqnarray}
The final result exactly agrees with the equivalent one in~\cite{Bartels:2001ge}:
\begin{eqnarray}
h^{(1)} (k_\perp)  &=& h^{(0)} (k_\perp) \frac{\alpha_s}{2\pi} \frac{\mathcal{P}_{gq}(z,\epsilon)}{\pi_\epsilon} \frac{1}{q_\perp^2 \Delta_\perp^2} 
\theta \left(z - e^{-\eta_b}\frac{\sqrt{q_\perp^2}}{p_a^+} \right)
\nonumber \\ &&
\qquad \qquad \qquad \qquad  \cdot 
\left[C_F z^2 {k'_\perp}^2 + N_c (1-z) {\Delta_\perp}\cdot {q_\perp} \right] 
.
\end{eqnarray}

As in the case of virtual correction it is now needed to subtract the
contribution from gluon production at central rapidities to construct
the complete differential cross section:
\begin{eqnarray}
   \parbox{4cm}{\center \includegraphics[height=1.3cm]{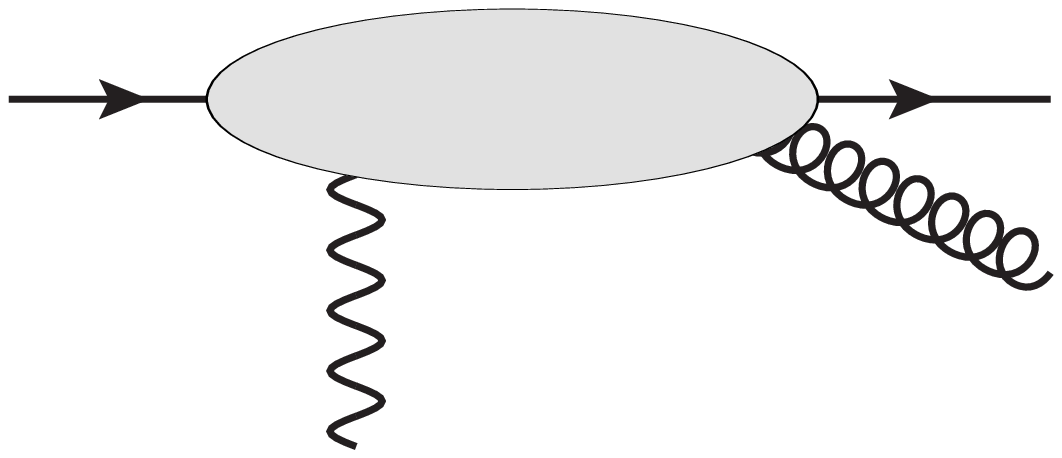}} &=&
\parbox{4cm}{\center \includegraphics[height=1.3cm]{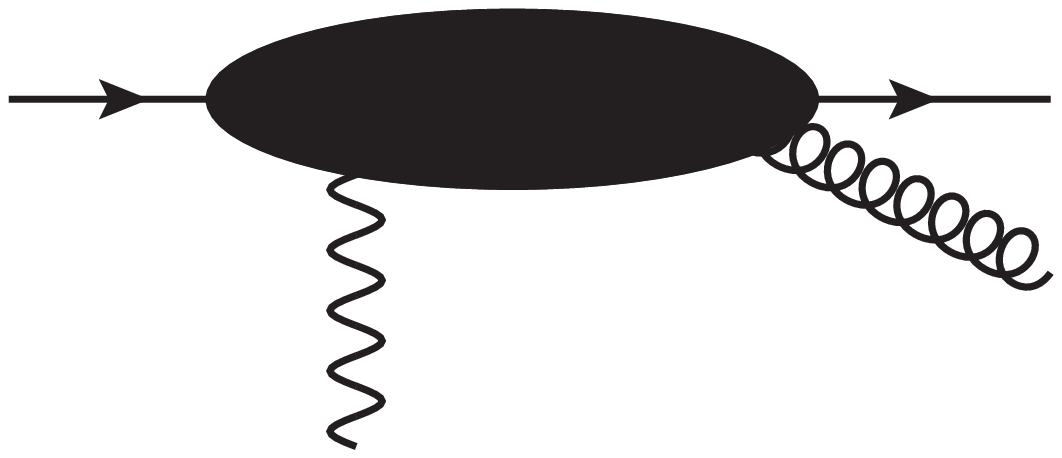}}  - \parbox{4cm}{\center \includegraphics[height=1.3cm]{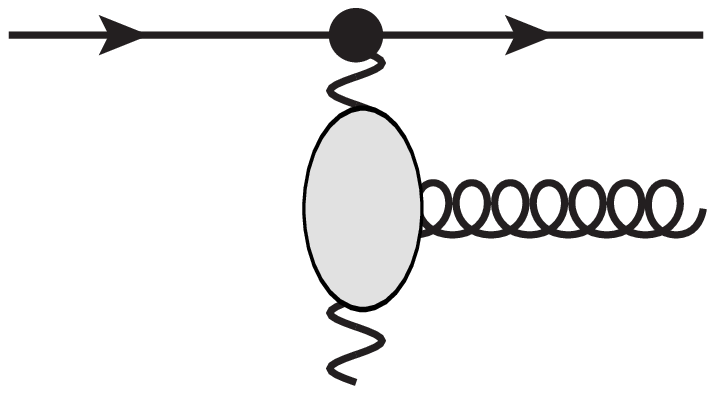}} \nonumber
\end{eqnarray}

In this way the quasi-elastic contribution to the exclusive differential cross-section is
\begin{eqnarray}
d \hat{\sigma}^{(qea)}_{ab} &=& h^{(0)}_a (k'_\perp) h^{(0)}_b(k_\perp)  
\mathcal{G}_{qqg}(k_\perp, q_\perp,z, \eta_a, \eta_b )  d^{2+2\epsilon} {q_\perp} d^{2+2\epsilon} {k_\perp} dz 
\end{eqnarray}
where
\begin{eqnarray}
\mathcal{G}_{qqg} \!\! &\equiv& \!\! \!\!
\frac{\alpha_s}{2\pi}  \left\{ \frac{\mathcal{P}_{gq}(z,\epsilon)}{\pi_\epsilon} 
\left[ \frac{ C_F z^2 k^{'2}_\perp}{q_\perp^2 \Delta_\perp^2}
 + \frac{N_c (1-z) {\Delta_\perp} \cdot {q_\perp}}{{q_\perp}^2 \Delta_\perp^2}
 - \frac{N_c}{z} \frac{1}{k^{'2}_\perp} \right] \right\} 
 \theta \left(z - e^{-\eta_b}\frac{\sqrt{q_\perp^2}}{p_a^+} \right) \nonumber\\
 &&\hspace{1cm}+\frac{N_c}{z} \frac{1}{k^{'2}_\perp}
\theta\left(z - e^{-\eta_b}\frac{\sqrt{q_\perp^2}}{p_a^+} \right)
\theta\left(z - e^{\eta_a}\frac{\sqrt{q_\perp^2}}{p_a^+} \right).
\end{eqnarray}
The complete exclusive differential cross-section is then given as the sum of central and quasi-elastic contributions:
\begin{eqnarray}
  d \hat{\sigma}_{ab} &=&  d \hat{\sigma}^{(c)}_{ab} +  d \hat{\sigma}^{(qea)}_{ab} +  d \hat{\sigma}^{(qeb)}_{ab}.
\end{eqnarray}

\section{Conclusions and Outlook}

In this letter an advanced application of Lipatov's effective action for the description of QCD processes 
at high energies is explained in detail. The calculation of the most interesting jet vertex for applications to 
hadronic phenomenology has been performed, finding precise agreement with previous results in the literature, 
obtained with more conventional approaches. A particular regularization of longitudinal divergencies has been 
proposed which allows for the determination of NLL matrix elements. Applications of this technique to other 
processes and, in particular, to the extraction of the gluon-initiated jet vertex and 
two-loop gluon Regge trajectory are being carried out in a parallel work. \\\\
{\bf Acknowledgements} Discussions with our collaborators J.~Bartels and L.~Lipatov are acknowledged. 
Research partially funded by European Comission (LHCPhenoNet PITN-GA-2010-2645649), Comunidad de Madrid (HEPHACOS ESP-1473) and German Academic Exchange Service (DAAD).

\end{document}